\documentclass[12pt]{article}
\def\be{\begin{equation}}
\def\ee{\end{equation}}
\def\ba{\begin{eqnarray}}
\def\ea{\end{eqnarray}}

\def\lb{\label}
\def\bb{\bibitem}
\def\dfrac{\displaystyle\frac}

\begin{document}



\begin{titlepage}
\date{January 26, 2024 }
\title{
\begin{flushright}\begin{small} LAPTH-007/24 \end{small} \end{flushright} \vspace{1cm}
Farewell to Cotton gravity}



\author{G\'erard Cl\'ement$^a$\thanks{Email: gclement@lapth.cnrs.fr} and
Khireddine Nouicer$^{b}$\thanks{Email: khnouicer@univ-jijel.dz} \\ \\
$^a$ {\small LAPTh, Universit\'e Savoie Mont Blanc, CNRS, F-74940  Annecy, France} \\
$^b$ {\small LPTh, Department of Physics, University of Jijel,} \\
{\small BP 98, Ouled Aissa, Jijel 18000, Algeria}}

\maketitle

\begin{abstract}
We reformulate our proof of the under-determination of Cotton gravity in terms of the Codazzi parametrization.
\end{abstract}

\end{titlepage}
\setcounter{page}{2}
In a recent paper \cite{Clement2023}, we showed that the theory of Cotton gravity\footnote{Not to be confused with the work by F.W. Cotton, ``A generalisation of the Einstein-Maxwell equations'', Eur. Phys. J. Plus 136 (2021) 162.} proposed by Harada \cite{Harada2021} is under-determined, the degree of arbitrariness in the solutions increasing with the degree of symmetry. However this critique of Cotton gravity was challenged by the authors of \cite{Sussman2024}, who argued that their alternative ``Codazzi parametrization'' formulation of Cotton gravity \cite{Mantica2023} was free from this defect. Actually, we did mention the Codazzi approach to Cotton gravity in our paper, stating that this approach being stritly equivalent to the original Harada formulation of Cotton gravity, the under-determination of the equations of Cotton gravity for highly symmetric configurations carries over to the under-determination of the Codazzi parametrization for the same configurations. While this should be obvious, we feel it necessary to reformulate and expand our argument in this short note.

First, let us correct several erroneous statements on our paper \cite{Clement2023} made in \cite{Sussman2024}. In the Introduction, we are alleged to have claimed that the shortcomings of Cotton gravity become more severe as the theory is applied to spacetimes with weaker symmetries, which is the exact opposite of our assessment. In subsection 2.1, the authors of \cite{Sussman2024} misrepresent our argument on anisotropic cosmologies as claiming that the contracted Bianchi identity restricts the set of vacuum solutions of Cotton gravity, which again is the exact opposite of our finding.

The field equations of Cotton gravity \cite{Harada2021} are
\be\lb{eqcott}
C_{\mu\nu\rho} = 8\pi G M_{\mu\nu\rho},
\ee
where
\be\lb{defcott}
C_{\mu\nu\rho} \equiv D_\mu R_{\nu\rho} - D_\nu R_{\mu\rho} - \dfrac16\left(g_{\nu\rho}D_\mu R - g_{\mu\rho}D_\nu R\right)
\ee
(with $R_{\mu\nu}$ the Ricci tensor for the metric $g_{\mu\nu}$, $R$ its trace, and $D_\mu$ the covariant derivative) is the Cotton tensor, and
the 3-tensor $M_{\mu\nu\rho}$ is related to the matter energy-momentum tensor $T_{\mu\nu}$ by
\be\lb{gam}
M_{\mu\nu\rho} \equiv D_\mu T_{\nu\rho} - D_\nu T_{\mu\rho} - \dfrac13\left(g_{\nu\rho}D_\mu T - g_{\mu\rho}D_\nu T\right).
\ee
Any solution of the Einstein equations with cosmological constant solves the equations of Cotton gravity. On the other hand, the space of solutions of Cotton gravity is clearly larger than in the case of general relativity. In brief, our critique of Cotton gravity in \cite{Clement2023} rested on the observation that the Bianchi identity for the Ricci tensor led to the constraint on the Cotton tensor
\be\lb{bianchi}
g^{\nu\rho} C_{\mu\nu\rho} \equiv 0
\ee
which, in the case of highly symmetric configurations, lowered the number of independent equations (\ref{eqcott}), and thus increased the degree of arbitrariness in the solutions.

Now for the Codazzi approach. Define a symmetric 2-tensor $C_{\mu\nu}$ of trace $C$ by the generalized Einstein equation
\be\lb{defcod}
R_{\mu\nu} - \dfrac12 R\,g_{\mu\nu} \equiv 8\pi G \left[T_{\mu\nu} + C_{\mu\nu} - C g_{\mu\nu}\right],
\ee
or equivalently
\be\lb{defcod2}
C_{\mu\nu} \equiv R_{\mu\nu} - \dfrac16 R\,g_{\mu\nu} - 8\pi G \left[T_{\mu\nu} - \dfrac13 T\,g_{\mu\nu}\right].
\ee
If the Cotton gravity equations (\ref{eqcott}) are satisfied, the tensor $C_{\mu\nu}$ defined by (\ref{defcod2}) fulfils the condition:
\be\lb{charcod}
E_{\mu\nu\rho} = 0,
\ee
where the 3-tensor $E_{\mu\nu\rho}$ is defined by
\be\lb{defE}
E_{\mu\nu\rho} \equiv D_\mu C_{\nu\rho} - D_\nu C_{\mu\rho}.
\ee
The condition (\ref{charcod}) on the tensor defined by (\ref{defE}) is the characteristic property of Codazzi tensors. Inserting the definition (\ref{defcod2}) in the definition (\ref{defE}) leads to the equivalent form
\be
E_{\mu\nu\rho} \equiv C_{\mu\nu\rho} - 8\pi GM_{\mu\nu\rho}.
\ee
Thus the Codazzi condition (\ref{charcod}) is equivalent to the fundamental equation (\ref{eqcott}) of Cotton gravity. For the sake of clarity we shall refer to the original formulation of Cotton gravity by equations (\ref{defcott}) and (\ref{eqcott}) as the Harada formulation, and to that by (\ref{defcod}) and (\ref{charcod}) as the Codazzi formulation. It should be clear that these are two equivalent formulations of the same theory.

Recalling that the components of the Cotton tensor $C_{\mu\nu\rho}$ are related by the algebraic identity (\ref{bianchi})
and that the matter tensor $M_{\mu\nu\rho}$ must obey the similar condition
\be
g^{\nu\rho}M_{\mu\nu\rho} = 0,
\ee
equivalent to the conservation law of the energy-momentum tensor, it follows that the components of the tensor $E_{\mu\nu\rho}$ are related by
\be\lb{constr}
g^{\nu\rho}E_{\mu\nu\rho} = 0,
\ee
independently of the Codazzi condition (\ref{charcod}). This constraint lowers the number of independent Codazzi equations, which for highly symmetric configurations becomes smaller than the number of independent metric functions, so that these cannot be determined completely.

Consider the example of static spherically symmetric configurations, which may be parameterized by
\be
ds^2 = - e^{2\nu(r)}\,dt^2 + e^{2\lambda(r)}\,dr^2 + r^2(dr^2 + \sin^2\theta\,d\varphi^2).
\ee
Owing to the spherical symmetry of both the metric and the matter source, the non-vanishing components of the tensor $C_{\mu\nu}$ defined by (\ref{defcod2}) are diagonal, the angular components being related by $C_{\varphi\varphi} = \sin^2\theta\, C_{\theta\theta}$. Clearly the radial component $C_{rr}$ do not contribute to the Codazzi condition (\ref{charcod}), so that there are only two non-trivial Codazzi equations $E_{trt} = 0$ and $E_{\theta r\theta} = 0$. However, these are not independent, because of the constraint (\ref{constr}), which reads here
\be
g^{tt}E_{trt} + 2\,g^{\theta\theta}E_{\theta r\theta} = 0.
\ee
It follows that there is only one Codazzi equation for two unknown metric functions of $r$, which therefore can be determined only up to one arbitrary function of $r$.

Our second example will be that of Friedman-Lema\^{\i}tre-Robertson-Walker (FLRW) cosmologies, parameterized by
\be\lb{FLRW}
ds^2 = - dt^2 + a^2(t)\bar{g}_{ij}dx^idx^j, \qquad  \bar{g}_{ij} = (1-kr^2)^{-1}\delta_{ij}
\ee
($i,j = 1,2,3$). Let us show by direct computation that the scale function $a(t)$ is not determined by the Codazzi condition. The non-vanishing components of the Ricci tensor are
\be
R_{00} =-3 \dfrac{\ddot{a}}a, \quad R_{ij} =  \left(a\ddot{a} + 2\dot{a}^2 + 2k\right)\bar{g}_{ij}
\ee
(where $x^0 = t$). In vacuum (${T_{\mu\nu} = 0}$) \footnote{The inclusion relation $\{T_{\mu\nu} = 0\} \subset \{M_{\mu\nu\rho} = 0\}$ is quite obvious.}, from (\ref{defcod2}),
\be
C_{00} = -\dfrac{2\ddot{a}}a + \dfrac{\dot{a}^2}{a^2} + \dfrac{k}{a^2}, \quad C_{ij} = \left(\dot{a}^2 + k\right)\bar{g}_{ij},
\ee
leading to
\be
E_{0 ij} = \dot{C}_{ij} - \dfrac12\left(\dot{g}_{ik}C^k_j + \dot{g}_{ij}C^0_0\right) \equiv 0,
\ee
for any function $a(t)$. This means that the tensor $C_{\mu\nu}$ is, in the absence of matter sources, a Codazzi tensor for the spacetime (\ref{FLRW}) whatever the form of the scale function $a(t)$, which is thus left completely arbitrary.

We have shown that there are cases where the defining equations (\ref{defcod}) and (\ref{charcod}) of the Codazzi formulation of Cotton gravity do not carry any information. Therefore, whatever the formulation, Harada or Codazzi, Cotton gravity does not qualify as a physical theory.

\end{document}